\documentclass[%
nofootinbib,
 amsmath,amssymb,
 aps,
 prd,
floatfix,notitlepage,
twocolumn, superscriptaddress,
]{revtex4-1}
\usepackage[normalem]{ulem}
\usepackage{slashed}
\usepackage{graphicx}
\usepackage{dcolumn}
\usepackage{bm}
\usepackage{hyperref}
\usepackage{orcidlink}
\usepackage{footnote}
\usepackage{hyperref}
\usepackage{bbold}
\usepackage{soul}

\begin{document}

\title{Chiral symmetry breaking and pion condensation in the early universe}

\author{Osvaldo {\sc Ferreira}\orcidlink{0000-0002-6711-8308}}
 \affiliation{
 Instituto de F\'\i sica, Universidade Federal do Rio de Janeiro,
 CEP 21941-972 Rio de Janeiro, RJ, Brazil 
}

\affiliation{Institut für Theoretische Physik, J. W. Goethe Universität,
 Max von Laue-Str. 1, 60438 Frankfurt am Main, Germany}

\author{Eduardo S. {\sc Fraga}\orcidlink{0000-0001-5340-156X}} 

\affiliation{
 Instituto de F\'\i sica, Universidade Federal do Rio de Janeiro,
 CEP 21941-972 Rio de Janeiro, RJ, Brazil 
}

\author{Maurício {\sc Hippert}\orcidlink{0000-0001-5802-3908}}
\affiliation{Centro Brasileiro de Pesquisas Físicas, Rua Dr. Xavier Sigaud 150, 
Rio de Janeiro, RJ, 22290-180, Brazil}

\affiliation{Instituto de Física, Universidade do Estado do Rio de Janeiro, Rua São Francisco Xavier, 524, Rio de Janeiro, RJ, 20550-013, Brazil}

\author{Jürgen {\sc Schaffner-Bielich}\orcidlink{0000-0002-0079-6841}}
\affiliation{Institut für Theoretische Physik, J. W. Goethe Universität,
 Max von Laue-Str. 1, 60438 Frankfurt am Main, Germany}


\date{\today}
\begin{abstract}
We determine the possible trajectories the universe may have followed in the QCD phase diagram during the QCD epoch. We focus on the roles of chiral symmetry breaking and pion condensation under high imbalances in lepton asymmetry. Adopting the quark-meson model as an effective description of QCD at finite temperature, charge and baryon chemical potentials we show that, for sufficiently large but physically motivated asymmetries, the universe may have entered the pion condensation phase through a first-order phase transition, followed by a second-order phase transition when exiting it. Such a first-order phase transition represents a new possible source of primordial gravitational waves during the QCD epoch. 

\end{abstract}

\maketitle


The origin of the matter-antimatter asymmetry remains a mystery despite many theoretical and observational efforts in the last decades \cite{Dine:2003ax, Buchmuller:2005eh}. This is partially due to the difficulty in probing the early stages of the cosmic evolution ($\approx 1$ s or less). In particular, the baryon asymmetry, given by the ratio between the baryon number density and the total entropy, $b=n_B/s$, can be precisely determined by CMB (\textit{Cosmic Microwave Background}) and BBN (\textit{Big Bang Nucleosynthesis}) observations to be $b=(8.7 \pm 0.06) \times 10^{-11}$ \cite{Planck:2018vyg}. However, the reason for this value remains unknown and many attempts to explain it predict that a large total primordial lepton asymmetry $l=\sum_\alpha n_{L_{\alpha}}/s$ ($\alpha = e, \mu, \tau$), with $n_{L_{\alpha}}$ the lepton number density, may have been present before BBN \cite{Buchmuller:2005eh, Flanz:1994yx, Davidson:2008bu, Barenboim:2016shh, Akita:2025zvq}. This quantity has been poorly constrained by observations and the bounds on the asymmetries for the individual leptons are even weaker. It has been shown in Refs. \cite{Domcke:2025lzg, Froustey:2021azz, Froustey:2024mgf, Barenboim:2016shh} using neutrino transport calculations, that one can have large individual primordial lepton asymmetries that will eventually become small when neutrino oscillations start at temperature $T\approx 10$ MeV. Within this scenario, one is naturally lead to investigating the consequences of such large lepton asymmetries on the physics of the standard model and asking what observational signatures might be generated. In this work, we explore how the QCD sector would be affected, and look in particular into the roles of chiral symmetry breaking and the possibility of pion condensation.

\begin{figure}[ht]
    \centering
    \includegraphics[height=6cm]{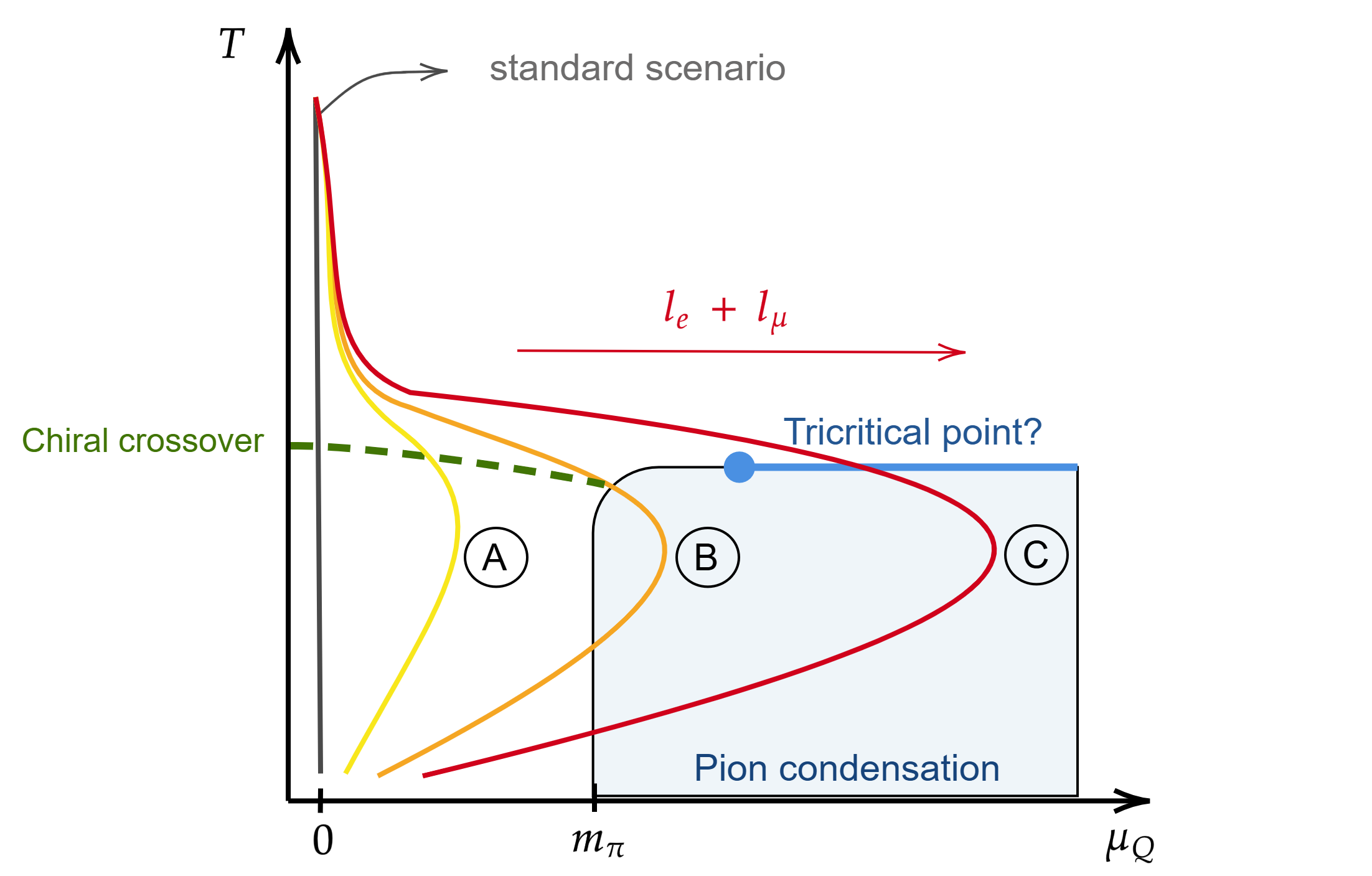}
    \caption{Pion condensation in the early universe. When baryons appear, charge neutrality can still be maintained at high $\mu_Q$. The universe can then reach phases such as pion condensation, depending on the imbalance in the individual lepton asymmetries quantified by $l_e+l_\mu$ (with $l_e + l_\mu + l_\tau =0$). The dark gray line is the standard trajectory that assumes no imbalance in the asymmetries and a total asymmetry of the order of $b$. In the figure three possibilities are illustrated: A) at small $l_e+l_\mu$ the trajectory does not reach the pion condensate; B) at intermediate values of $l_e+l_\mu$ the universe enters and leaves the pion condensate through second-order PTs; C) at high $l_e+l_\mu$ the universe enters the pion condensed phase through a first-order PT and exits it through a second order PT.}
    \label{fig:cartoon_pion_condensation}
\end{figure}

QCD at finite temperature and chemical potentials has a rich phenomenology that gives rise to a variety of complex phase diagrams \cite{Guenther:2020jwe}. For example, it is known from Lattice QCD that, if high charge chemical potentials ($\mu_Q \geq m_\pi$, with $m_\pi$ the pion mass) are reached at around $140$ MeV, pion condensation takes place \cite{Brandt:2017oyy, Son:2000xc, Son:2000by}. It is usually assumed that in the early universe $l$ and $b$ are of the same order, which would make both charge ($\mu_Q$) and baryon ($\mu_B$) chemical potentials very small and result in the universe following a simple trajectory along the temperature axis of the QCD phase diagram (standard scenario in Fig. \ref{fig:cartoon_pion_condensation}). However, there is also the possibility that an imbalance between lepton asymmetries or even a nonzero total lepton asymmetry was present during the QCD epoch, which would have lead to high values of charge and baryon chemical potentials \cite{Wygas2018otj, Vovchenko2020crk}. 
Depending on the lepton asymmetries, the universe may have taken different paths in the QCD phase diagram, often referred to as \emph{cosmic trajectories}, possibly crossing phase transitions as it cooled down.

 The high values of charge chemical potential reached when high lepton asymmetries are present, naturally lead to the prediction of pion condensation, as first pointed out in \cite{Wygas2018otj, Middeldorf-Wygas:2020glx, abuki2009}. The conditions for pion condensation at the QCD epoch were later analyzed in detail by Ref. \cite{Vovchenko2020crk}, where a pion-condensed phase at large $\mu_Q$ was explicitly introduced using an effective mass model matched to lattice data. In this work, we reexamine this possibility within the two-flavor quark-meson (QM) model. Despite its simplicity, this model correctly exhibits the approximate chiral symmetry breaking pattern $SU_L(2)\times SU_R (2) \rightarrow SU_V (2)$ and its relation with the pion mass \cite{Koch:1997ei}, and has been used in a variety of investigations of the QCD phase diagram \cite{kamikado2013a, Skokov:2010sf, andersen2009, Palhares:2009tf, Khan:2011wai, He:2005nk, ebert2006}. 

Therefore, this approach provides a complementary analysis, addressing points not discussed in Ref. \cite{Vovchenko2020crk}.\footnote{Ref. \cite{Cao:2021gfk} uses a PNJL model to discuss the cosmic trajectories. There, the order parameters are the Polyakov loop and the pion condensate. Although the assumptions of both treatments are similar, the discussions take different directions regarding which information to extract from the phase diagrams and thermodynamic quantities.} Namely, within the QM model: (i) we discuss both the chiral phase transition and the pion condensation phase; (ii) we evaluate the corresponding condensates along the cosmic trajectories by solving the conservation equations for the early universe while simultaneously minimizing the free energy of the system with respect to the condensates, i.e., solving the gap equations; and (iii) we discuss the cosmic trajectories along the QM model phase diagram, which is different from the effective mass model, exhibiting a tricritical point at high $\mu_Q$ and high $T$ (see Fig. \ref{fig:cartoon_pion_condensation}). Since a full determination of the QCD phase diagram at finite $T$, $\mu_B$ and $\mu_Q$ is lacking, and a possible tricritical point at high values of $\mu_Q$ ($\gtrsim 1.5 m_\pi$) has not been investigated on the Lattice ($\mu_B=0$), we consider this scenario worth investigating given its possible phenomenological consequences. For instance, if the universe went through phase transitions as it cooled down, this may have left imprints on a pre-existing gravitational wave (GW) background, or even generated additional gravitational waves if the universe went through a first-order phase transition (PT).

This work is organized as follows. In Section \ref{section: model} we introduce and discuss the quark-meson model that will be employed in the description of the QCD sector. In Section \ref{section: EU_conditions} we discuss the conditions that were present in the early universe and that determine the path the universe should follow through the phase diagram. In Section \ref{section: results} we discuss the phase diagram of the QM model, the results for the cosmic trajectories and for the behavior of the condensates, as well as what these results imply for the phase transitions that the universe may undergo for given lepton asymmetries imbalances. In Section \ref{section: discussion} we discuss the results and point to some possible phenomenological consequences, like the generation of GWs. Section \ref{section: conclusions} summarizes the main conclusions and outlines perspectives for future work.

\section{The model}\label{section: model}

\subsection{Quark-Meson model}

The quark-meson model was shown to be very suitable for the study of the chiral transition long ago \cite{lee-book}. As argued originally in Ref. \cite{Pisarski:1983ms}, QCD with two flavors of massless quarks belongs to the same universality class as the $O(4)$ linear sigma model, exhibiting the same qualitative behavior at criticality. This model has also the good feature of being renormalizable \cite{Lee:1968da}, even though this is not necessary for an effective model, and reproduces correctly the phenomenology of strong interactions at low energies. Since its proposal \cite{Gell-Mann:1960mvl}, it has been investigated in different contexts, and very early used to study the chiral transition of thermal QCD \cite{Baym:1977qb,Bochkarev:1995gi,Bilic:1997sh,Petropoulos:1998gt,Scavenius:1999zc,Roder:2003uz,Scavenius:2000qd,Scavenius:2000bb,Fraga:2004hp,Koide:2004yn,Sasaki:2007qh,Marko:2010cd,Nahrgang:2011mv}, and also at finite density \cite{Palhares:2010be,Kahara:2010wh,Kroff:2014qxa}.

At finite quark chemical potential $\mu_q=\frac{\mu_B}{3}$, $\mu_B$ being the baryon chemical potential, and isospin $\mu_I=\frac{\mu_Q}{2}$, we can write the Lagrangian for the Quark-Meson model $\mathcal{L}_{QM}$ as \cite{kamikado2013a, Strodthoff:2011tz, andersen2009, HippertPhD}
\begin{equation}
\begin{aligned}
\mathcal{L}_{\mathrm{QM}}= &\mathcal{L}_M + \bar{\psi}\left( \slashed{\partial} +\gamma_0 \hat{\mu}+i g\left(\sigma+i \gamma_5 \vec{\tau}\cdot \vec{\pi}\right)\right) \psi+ \\
& -2 \mu_I\left(\pi_1\left(\partial_0 \pi_2\right)-\pi_2\left(\partial_0 \pi_1\right)\right)-2 \mu_I^2\left(\pi_1^2+\pi_2^2\right)
\end{aligned}
\end{equation}
where
\begin{equation}
    \hat{\mu}=\left(\begin{array}{cc}
\mu_q+\mu_I & 0 \\
0 & \mu_q-\mu_I
\end{array}\right)
\end{equation}
and 
\begin{equation}\label{eq: meson_lagrangian}
    \mathcal{L}_{M} = \frac{1}{2} \partial_\mu \sigma \partial^\mu \sigma
    + \frac{1}{2} \partial_\mu \vec{\pi} \cdot \partial^\mu \vec{\pi}
    + U\left(\sigma, \vec{\pi}, \mu_I\right),
\end{equation}
with
\begin{equation}
    U\left(\sigma, \vec{\pi}, \mu_I\right) = \frac{\lambda}{4} \left(\sigma^2 + \pi^2 - v^2\right)^2
    - h \sigma - 2 \mu_I^2 \pi^2.
\end{equation}
Here, the scalar field $\sigma$ corresponds to the sigma meson and the components of $\vec{\pi}$ correspond to the pion fields. We adopt the decomposition  $\sigma = \sigma_0 + \bar{\sigma}$  and $\vec{\pi}= (\pi_0 + \pi_{\|}, \pi_{\perp}, \pi_3)$, where $\sigma_0$ and $\pi_0$ are the mean values (or condensates) of the sigma and pion fields and $\bar{\sigma}$ and $(\pi_{\|}, \pi_{\perp}, \pi_3)$ their corresponding thermal fluctuations. The $N_f=2$ fermion fields $\psi=(u,d)$ represent the up and down constituent quarks. Notice that the self-interacting potential in the mesonic Lagrangian given by Eq. \eqref{eq: meson_lagrangian} exhibits a spontaneous symmetry breaking pattern $SU_L(2)\times SU_R (2) \rightarrow SU_V (2) $. Moreover, the term $h\sigma$ corresponds to an explicit symmetry breaking, which is a consequence of a finite current quark mass. 

Following the standard approach in the mean-field approximation, we neglect thermal fluctuations of mesons and replace the meson fields by their mean values, arriving at the following free energy:
\begin{equation}\label{eq: QM_thermodynamical_potential}
\begin{aligned}
\Omega_{\text{QM}} = &\Omega_{\text{vac}} - 2 N_c T \int \frac{d^3 p}{(2 \pi)^3} \times \\
&\bigg[ \ln \left(1+e^{-(E_{\Delta}^{-}+\mu_q)/T}\right) + \ln \left(1+e^{-(E_{\Delta}^{-}-\mu_q)/T}\right) \\
&+ \ln \left(1+e^{-(E_{\Delta}^{+}+\mu_q)/T}\right) + \ln \left(1+e^{-(E_{\Delta}^{+}-\mu_q)/T}\right) \bigg] \\
&+ U(\sigma_0, \pi_0, \mu_I)\,,
\end{aligned}    
\end{equation}
where the vacuum term has the form
\begin{equation}
    \Omega_{vac}= -2 N_c \int \frac{d^3 p}{(2 \pi)^3}\left(E_{\Delta}^{+}+E_{\Delta}^{-}\right)
\end{equation}
with
\begin{equation}
    E_{\Delta}^{\pm}= \sqrt{\left(E_p \pm \mu_I \right)^2 + \Delta^2}
\end{equation}
Here, $E_p= \sqrt{p^2 + m^2}$, $m=g\sigma_0$ and $\Delta=g \pi_0$. For simplicity, in the following discussion we neglect the vacuum term in Eq. \eqref{eq: QM_thermodynamical_potential}, using the so-called no-sea approximation (for a discussion on the relevance of vacuum terms see Refs. \cite{Mocsy:2004ab,Palhares:2008yq,Fraga:2009pi,Boomsma:2009eh,Palhares:2010be,Mizher:2010zb,Skokov:2010sf,Strodthoff:2011tz,kamikado2013a,Haber:2014ula}). 
The free parameters of the model are the coupling constant $g$, the self-interacting coupling $\lambda$, the constant in the explicit symmetry breaking term $h$, and the vacuum expectation value at $h=0$, $v$. We fix these parameters by requiring that the model should reproduce vacuum values of the pion decay constant $f_\pi=92$ MeV, the pion mass $m_\pi=138$ MeV and $m_\sigma =700$ MeV. The value of $m_\sigma$ was chosen in order to make the chiral phase transition occur at $T \approx 160$ MeV at $\mu_B=0$ and $\mu_Q=0$, in agreement with Lattice simulations \cite{Brandt:2017oyy}.

\section{Early universe conditions}\label{section: EU_conditions}

In the early stages of cosmic evolution ($\lesssim 1s$) the rate of interactions between particles is much larger than the Hubble rate, $\Gamma \gg H$. This allows us to describe the universe as a plasma of relativistic particles in thermal equilibrium that conserves baryon number $B$, charge $Q$ and lepton number $L$. As the universe cools, heavier particles will decouple from thermal equilibrium, changing the particle content of the cosmic plasma. This particle content in turn affects the expansion of the universe through the Friedmann equations \cite{Weinberg:2008zzc}. Therefore, knowing how matter behaves is essential for a good description of the evolution of the universe at its early stages. The cosmic trajectories determined in this paper bring information about the behavior of matter during the QCD epoch, considering in particular the possibility of phase transitions when high charge chemical potentials are allowed.

To describe the path that the universe followed during the QCD epoch we must take into account some observational constraints. The asymmetry $x$ associated to a given conserved quantity $X$ is a measure of the imbalance in the number of particles that carry positive and negative values of this quantity, normalized by the total entropy density. Measurements from the CMB can place strong constraints on the possible electrical charge asymmetry in the universe. For instance, the authors of Ref. \cite{Caprini:2003gz} used the isotropy of CMB to state that $q_{e-p}< 10^{-26}e$ in the case of a uniform distribution of charge, showing that observations are consistent with a charge neutral universe. On the other hand, the value for the baryon asymmetry $b$ can be extracted from BBN and CMB observations and is inferred to be $b=(8.7 \pm 0.06) \times 10^{-11}$ \cite{Planck:2018vyg}. The presence of lepton asymmetries in the universe after neutrino decoupling\footnote{Notice that after the baryon and lepton decoupling, charge neutrality implies that, since the baryon number density is measured to be small, the lepton number density of the charged leptons should also be small. Then, if primordial lepton asymmetries survived to the present day, they would be hidden in the neutrino sector.} can affect at least two important quantities that are relevant for the CMB spectrum and for BBN. The first effect is on the proton-to-neutron ratio $\frac{n_{n}}{n_{p}}$ via the electron-neutrino asymmetry; the second is on the effective number of neutrino degrees of freedom $N_{\mathrm{eff}}$. This can be used to constrain the possible values of $l$. For instance, the authors of Ref. \cite{Oldengott:2017tzj} find $l<10^{-2}$, but in Ref. \cite{Burns:2022hkq} its is argued that it could be larger depending on the details of the neutrino sector. In what follows we assume $l=0$ for simplicity, since the imbalance in individual asymmetries is sufficient to generate pion condensation.

Following an approach similar  to Refs. \cite{schwarz2009, Wygas2018otj, Vovchenko2020crk, gaoFunctional2022, hajkarim2019a}, we associate a chemical potential and a conservation equation to each of the charges that are conserved during the cosmic evolution that follows baryogenesis and leptogenesis \cite{Cline:2006ts}, i.e., the baryon number $B$, the electric charge $Q$ and the lepton number density $L_{\alpha}$ for each lepton flavor $\alpha=e, \mu, \tau$. Then, for a given temperature $T$, the equations 

\begin{equation}\label{eq: baryon_asymmetry}
\frac{n_B\left(T, \mu_B, \mu_Q\right)}{s\left(T, \mu_B, \mu_Q,\left\{\mu_\alpha\right\}\right)}  =b,
\end{equation}
\begin{equation}\label{eq: charge_asymmetry}
    \frac{n_Q\left(T, \mu_B, \mu_Q,\left\{\mu_\alpha\right\}\right)}{s\left(T, \mu_B, \mu_Q,\left\{\mu_\alpha\right\}\right)}  =0,
\end{equation}
\begin{equation}\label{eq: lepton_asymmetry}
    \frac{n_{L_\alpha}\left(T, \mu_Q,\left\{\mu_\alpha\right\}\right)}{s\left(T, \mu_B, \mu_Q,\left\{\mu_\alpha\right\}\right)} =l_\alpha, \quad \alpha \in e, \mu, \tau .
\end{equation}
define the point $(\mu_B, \mu_Q,\left\{\mu_\alpha\right\})$ in which the universe is located in this space. The cosmic trajectory of the universe can then be found by solving these equations in a range of temperatures. Here $n_B$ is the baryon number density and $n_Q$ is the charge number density. The lepton number density is given by $n_{L_\alpha}=n_{\alpha} + n_{\nu_{\alpha}}$, with $n_{\alpha}$ corresponding to the charged lepton number density, and the number density of its corresponding neutrino being given by $n_{\nu_{\alpha}}$. These number densities are divided by the total entropy $s$. Equations \eqref{eq: baryon_asymmetry}-\eqref{eq: lepton_asymmetry} define what we will from now on refer to as baryon asymmetry $b$, charge asymmetry $q=0$ and lepton asymmetry $l_\alpha$, respectively.

Notice that we can rewrite the particle chemical potentials in terms of the charge ($\mu_{Q}$), baryon ($\mu_{B}$) and lepton number ($\mu_{L_{\alpha}}$) chemical potentials:
\begin{equation}
\begin{aligned}
    \mu_u & =\frac{2}{3} \mu_Q+\frac{1}{3} \mu_B, \\
    \mu_d & =-\frac{1}{3} \mu_Q+\frac{1}{3} \mu_B, \\
    \mu_\alpha & =\mu_{L_\alpha}-\mu_Q, \\
    \mu_{\nu_\alpha} & =\mu_{L_\alpha} \, ,
\end{aligned}
\end{equation}
with $\mu_{\nu_{\alpha}}$ ($\alpha=e, \mu, \tau$) the chemical potentials of the neutrinos. In our model $\mu_q=\frac{1}{2}(\mu_u + \mu_d)=\frac{\mu_B}{3}$ which leads to $\mu_q=\frac{1}{3}\mu_B + \frac{1}{6}\mu_Q$.
To determine the cosmic trajectories we must minimize the quark-meson potential \eqref{eq: QM_thermodynamical_potential} with respect to the condensates of the model, $(\pi_0, \sigma_0)$, while imposing equations \eqref{eq: baryon_asymmetry}-\eqref{eq: lepton_asymmetry} as constraints. In the following discussion, based on the observation  that the occurrence of pion condensation depends essentially on the value of $l_e + l_\mu$ \cite{Vovchenko2020crk}, we will vary this quantity within the range  $0.01-0.4$. These are typical values found in models that predict generation of primordial asymmetries \cite{Dine:2003ax, Affleck:1984fy, schwarz2009, Casas:1997gx, McDonald:1999in, Abazajian:2004aj}

The total pressure, entropy and energy density that are used during the calculations are given by the sum of the contributions from the QCD sector, described by the QM model, with contributions from leptons and photons, under the constraints discussed in this section. The leptons will be treated as free massive relativistic particles and the photons as free massless relativistic particles.

\section{Results}\label{section: results}

\subsection{Phase diagram}
\begin{figure*}[htbp]
  \centering
  \begin{minipage}[t]{0.48\textwidth}
    \centering
    \includegraphics[height=4.9cm]{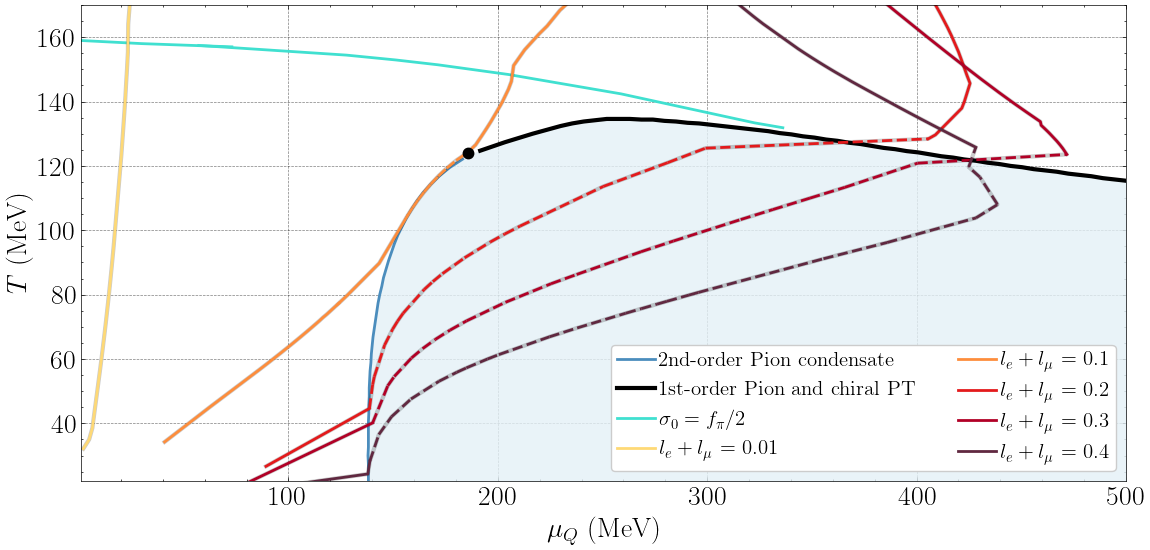}
    \label{fig: CT_in_the_phase_diagram}
  \end{minipage}
  \hfill
  \begin{minipage}[t]{0.48\textwidth}
    \centering 
    \includegraphics[height=4.9cm]{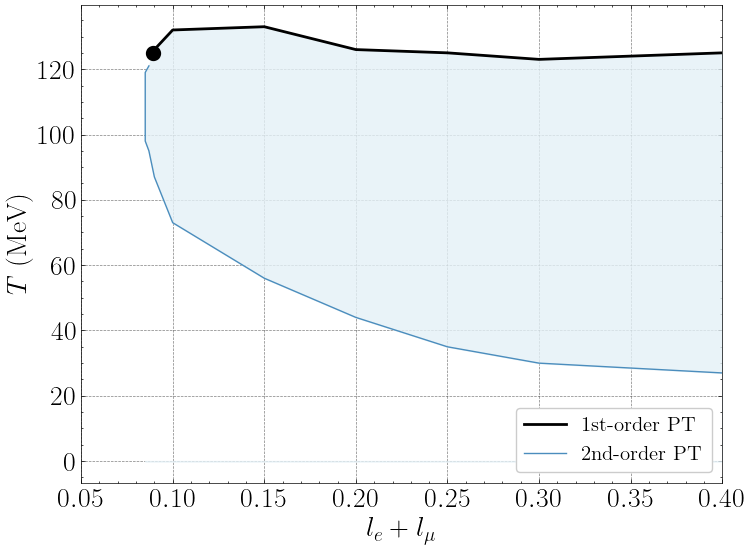}
    \label{fig: Pion condensation region}
  \end{minipage}
    \caption{Left: Cosmic trajectories in the $T$ vs $\mu_{Q}$ plane for different values of $|l_e + l_\mu|$. For illustration, the background shows the phase diagram of the QM model, at fixed $\mu_B=100$ MeV. The blue shaded region corresponds to the pion condensation phase. The dashed segments indicates that the trajectories have entered the pion condensation phase. Notice that $\mu_B$ varies along the cosmic trajectories. Right: Region in the $T$ vs $l_e + l_\mu$ phase diagram in which the pion condensation can be formed.}
    \label{fig: CT_and_pion_condensation}
\end{figure*}

In Fig. \ref{fig: CT_and_pion_condensation}, we show the phase diagram of the QM model in the $T$ vs $\mu_Q$ plane at $\mu_B = 100$ MeV. In the plot we see the phase boundary for pion condensation. The transition to this phase is of second order at low temperatures, but becomes first-order at a tricritical point located at $T_{tc}\approx 124$ MeV and $\mu_{Qc} \approx 186$ MeV ($\mu_I \approx 93$ MeV).
We also plot a line for half the value of the chiral condensate, \textit{i.e.}, $\frac{f_{\pi}}{2}$, until this line meets the pion condensation boundary, which we take as an estimate for the chiral crossover line (as done in Refs. \cite{kamikado2013a, Strodthoff:2011tz}). Notice that, after the tricritical point, the chiral condensate also exhibits signs of a first-order PT on the boundary (this causes the jumps observed in Fig. \ref{fig:Condensates_vs_T}.

\subsubsection{Comparison to other effective models} \label{subsection: other models}

The phase diagram exhibited in Fig. \ref{fig: CT_and_pion_condensation} is dependent on the underlying assumptions about the model, namely the choice of the QM model, the use of the mean-field approximation, and the disregard of vacuum terms. Since our goal in this work is to present a qualitative picture of possible scenarios at the QCD epoch when high lepton asymmetries, and therefore high charge chemical potentials are present, these simplifying assumptions are a convenient choice. It is nonetheless relevant to comment on how the model discussed in Section \ref{section: model} compares to other models in the literature. We focus, in particular, on the presence of a tricritical point in the pion condensation boundary in the $T$ vs $\mu_Q$ ($\mu_B=0$) plane, which is the most relevant feature for the present discussion. 

In the mean-field Nambu–Jona-Lasinio (NJL) model the tricritical point is also present \cite{Zhang:2006gu}, but at higher $\mu_Q$. And this remains true in the Polyakov-loop extended Nambu–Jona-Lasinio (PNJL) model, as discussed in \cite{Sasaki:2010jz}. Notice that Refs. \cite{andersen2009, He:2005nk,  Lopes:2021tro} do not see such tricritical point, but these investigations usually report on the phase diagram for $\mu_Q \lesssim 2m_{\pi}$. A tricritical point is also observed in the holographic model discussed in Ref. \cite{Kovensky:2024oqx}, even though the authors find a first-order chiral phase transition at low $\mu_Q$, in disagreement with lattice results. In the case of the QM model, depending on how mesonic fluctuations are included, a tricritical point may be present or not (see discussions in Refs. \cite{kamikado2013a, Strodthoff:2011tz}). 

One would not expect that a Polyakov-loop extension of the QM model would bring significant qualitative changes to the phase diagram of Fig.~\ref{fig: CT_and_pion_condensation}~ since, for the values of $\mu_B$ along the trajectories considered here (usually $\mu_B \lesssim 250$ MeV), the deconfinement transition line is expected to be a crossover and to be very close to the chiral transition line (see discussions in Refs. \cite{Adhikari:2018cea, Ueda:2013sia}).

Finally, the treatment of the vacuum terms may change the QM model phase diagram depicted in Fig. \ref{fig: CT_and_pion_condensation}. For instance, the position of the tricritical point has been reported to be dependent on the UV cutoff, with higher values for the cutoff scale leading to a tricritical point at higher $\mu_Q$ \cite{kamikado2013a, Strodthoff:2011tz}, and such tricritical point was not found within a renormalization-group invariant mean-field formulation \cite{Brandt:2025tkg} at the values of $\mu_Q$ investigated.

\subsubsection{Comparison to the Lattice}

Although our model considers $\mu_B \neq 0$, it is useful to mention what is known from Lattice QCD from which information at $\mu_B=0$ and finite isospin is available (for a detailed discussion see Ref. \cite{Brandt:2017oyy}). Let us make a brief comparison with some of the findings of ref. \cite{Brandt:2017oyy} (for other related results from Lattice QCD see also Refs. \cite{Abbott:2023coj, Detmold:2012wc, Mitra:2024czm, Kogut:2002tm, Kogut:2002zg}):
\begin{enumerate}
    \item At around $T \approx 120$ MeV Lattice simulations show that the behavior of the pion condensate indicates a second-order transition occurring at $\mu_{Qc}=2\mu_{Ic}= m_{\pi}$.%
\footnote{Notice that the plots in Ref. \cite{Brandt:2017oyy} follow the convention of dividing the isospin chemical potential by 2, so that there $\mu_{Ic}=\frac{m_{\pi}}{2}$.} 
This is consistent with our model, although the shape of the boundary of the pion condensation phase becomes comparatively more rounded at higher temperatures (compare the phase diagram in Fig.~\ref{fig: CT_and_pion_condensation} to Fig. 10 of ref. \cite{Brandt:2017oyy}).
    \item The pion condensation phase boundary meets the chiral phase transition line at a pseudo-tricritical point located at $T_{pt} \approx 151$ MeV \footnote{On the Lattice the critical temperature for the chiral crossover is defined to be located in the inflection point of the chiral condensate as a function of temperature. This definition is different from what is used in Fig. ~\ref{fig: CT_and_pion_condensation}, which exhibits the half-value of the chiral condensates. Nevertheless, we checked that for $\mu_Q < m_{\pi}$ the two definitions produce similar results.}. Up to this pseudo-tricritical point, the chiral phase transition is confirmed to be a crossover. We also observe a pseudo-tricritical point, but the chiral crossover meets the pion condensation phase boundary at higher values of isospin. Moreover, on the Lattice the chiral phase transition line is aligned with the upper part of the pion condensation phase boundary, which is not realized in our model.
    \item  For a fixed isospin value of $\mu_I\approx 104.5 $ MeV, Lattice QCD results show the boundary of the pion condensation phase to be located at $T \approx 158$ MeV. At this point, the condensate drops to zero rather smoothly, indicating a second-order phase transition. Besides, the chiral condensate also drops considerably, but remains continuous and smooth, which is consistent with a crossover. As mentioned previously, in the QM model there is a tricritical point in which the second-order phase transition of the pion condensation becomes a first-order phase transition (a similar behavior has been reported in holographic models \cite{Kovensky:2024oqx}). Although this last feature is not observed on the Lattice at this value of $\mu_Q$, to our knowledge the possibility of such a tricritical point at higher $\mu_Q$ has not been excluded. We thus decide to take the possibility of this tricritical point seriously in the following discussion, which may hopefully serve as a motivation for further investigations on Lattice QCD.
\end{enumerate}

Despite some differences with respect to the Lattice ($\mu_B=0$), the quark-meson model can describe many essential features of the QCD phase diagram at finite isospin. Additionally, recent developments have shown that extended versions of the quark-meson model can achieve remarkable agreement with the Lattice and may be incorporated into a future update of this analysis \cite{Brandt:2025tkg}. 

Within our framework, the main effect of a finite $\mu_B$ is to bring the phase boundary of the pion condensation at $\mu_Q > m_\pi$ to lower values of the temperature. Furthermore, it causes the tricritical point to occur at lower values of $\mu_Q$ \cite{kamikado2013a}. Then, if Lattice simulations find a tricritical point at high $\mu_Q$ for $\mu_B=0$, a finite $\mu_B$ could in principle shift it towards lower values of $\mu_Q$.

\subsection{Cosmic trajectories in the phase diagram}

In Fig.~\ref{fig: CT_and_pion_condensation} we show our results for the cosmic trajectories associated with different values of $|l_e + l_\mu|$, which is a measure of the imbalance between the individual lepton asymmetries. In all the trajectories we make the conservative assumption that $l_e + l_\mu + l_\tau =0$, i.e., the total lepton asymmetry vanishes. The dashed segments correspond to the part of the trajectories in which the universe is on the pion condensation phase.

Our results are consistent with those of Ref. \cite{Vovchenko2020crk}: once the threshold $|l_e + l_\mu| \gtrsim 0.1$ is reached, the cosmic trajectories enter the pion condensation phase. However, in our calculations slightly smaller values ($\sim 0.085$) can still generate pion condensation for a small range of temperatures. Another important difference is that the quark meson-model predicts a first-order PT for $|l_e + l_\mu| > 0.1$ when the universe enters the pion condensation phase and a second-order PT when it leaves it (case (C) in the cartoon depicted in Fig. \ref{fig:cartoon_pion_condensation}).
From the plot on the right-hand side of Fig. \ref{fig: CT_and_pion_condensation}, in which we exhibit the region in the $T$ vs $|l_e + l_\mu|$ plane where pion condensation occurs, we see that this corresponds to most of the trajectories that enter the pion condensation phase. The exceptions are in a small range of $|l_e + l_\mu|$ right before $|l_e + l_\mu| = 0.1$, for which both transitions are second-order (case (B) in Fig. \ref{fig:cartoon_pion_condensation}). However, in a more quantitatively precise analysis, we expect the region in which both transitions are second order to be larger, since the inclusion of vacuum terms usually pushes the tricritical point to higher $\mu_Q$ values, as discussed in Section \ref{subsection: other models}.

\begin{figure}[htbp]
  \centering
  \begin{minipage}[t]{0.48\textwidth}
    \vspace{0pt}
    \centering
        \includegraphics[height=5cm]{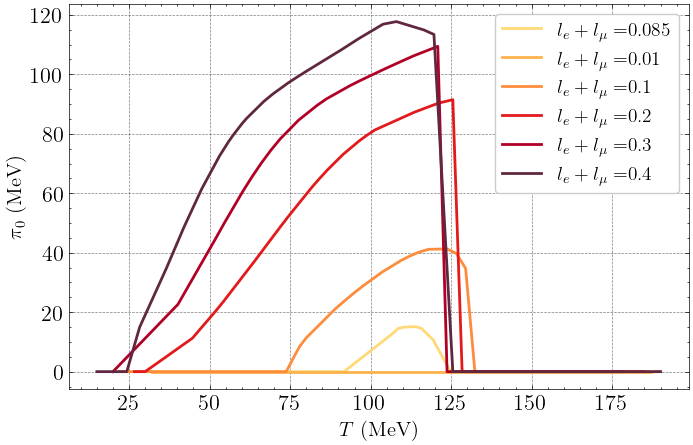}
        \label{fig:Pion_Condensate_vs_T}
  \end{minipage}
  \hfill
  \begin{minipage}[t]{0.48\textwidth}
    \vspace{0pt}
    \centering
        \includegraphics[height=5cm]{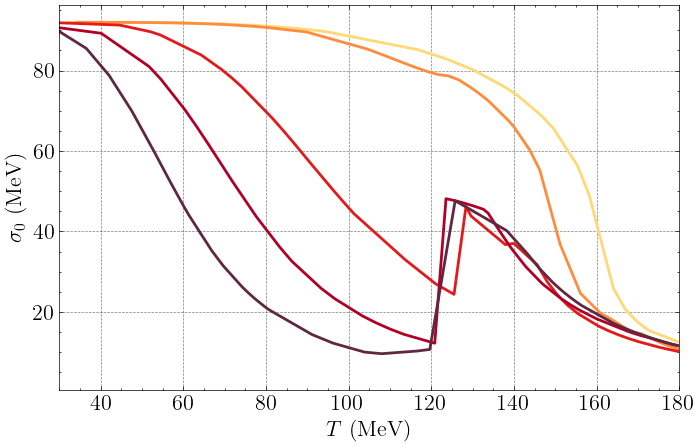}
        \label{fig:Chiral_Condensate_vs_T}
  \end{minipage}
  
    \caption{Condensate values along cosmic trajectories as a function of temperature. Pion condensate (left) and chiral condensate (right). The first-order phase transitions along the cosmic trajectories cause discontinuities in the condensates.}
    \label{fig:Condensates_vs_T}
\end{figure}

The occurrence of the transitions can be clearly seen in the behavior of the order parameters, i.e, the condensates along the trajectories, which are shown in Fig. \ref{fig:Condensates_vs_T}. We see that, for $|l_e + l_\mu| = 0.01$, the chiral condensate varies smoothly and shows an inflection point at $T\approx 160$ MeV, while the pion condensate is zero throughout the entire range of temperatures. This indicates that the cosmic trajectory goes through the chiral crossover and does not enter the pion condensate. For $|l_e + l_\mu| \geq 0.1$ the chiral condensate starts to exhibit a bump that turns into a jump for $|l_e + l_\mu| \geq 0.2$, which indicates that the trajectories go through the chiral transition which has turned into a first-order PT. Moreover, we see from Fig. \ref{fig:Condensates_vs_T} that, as $|l_e + l_\mu|$ increases, the first-order jump in the pion condensate becomes larger.   

The transition line also affects the shape of the different trajectories in the phase diagram on the left-hand side of Fig.~\ref{fig: CT_and_pion_condensation}. 
When crossing the first-order phase transition, cosmic trajectories move discontinuously along the transition line, in the direction of the tricritical point, due to the entropy and particle number densities being discontinuous across that line. 
This leads to the focusing of trajectories towards the tricritical point, as has been previously considered in the context of heavy-ion collisions 
\cite{Stephanov:1998dy,Stephanov:1999zu,Nonaka:2004pg,Stephanov:2004wx}.
The increase of susceptibilities around the second-order transition may also deform the cosmic trajectories, causing them to move preferentially along the transition line, similarly to what was noticed in \cite{Dore:2022qyz}.

\begin{figure}[htbp]
    \centering
  \begin{minipage}[t]{0.48\textwidth}
       \vspace{0pt}
       \centering
        \includegraphics[width=\linewidth]{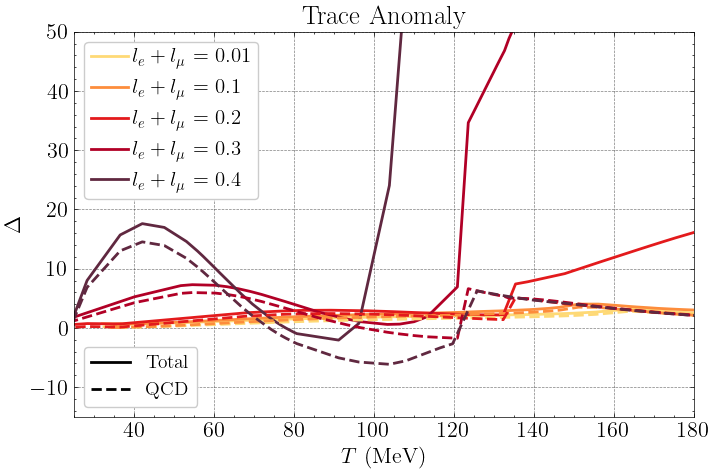}
        \label{fig:TraceAnomaly_vs_T}
  \end{minipage}
  \hfill
  \begin{minipage}[t]{0.48\textwidth}
    \vspace{0pt}
    \centering
        \includegraphics[width=\linewidth]{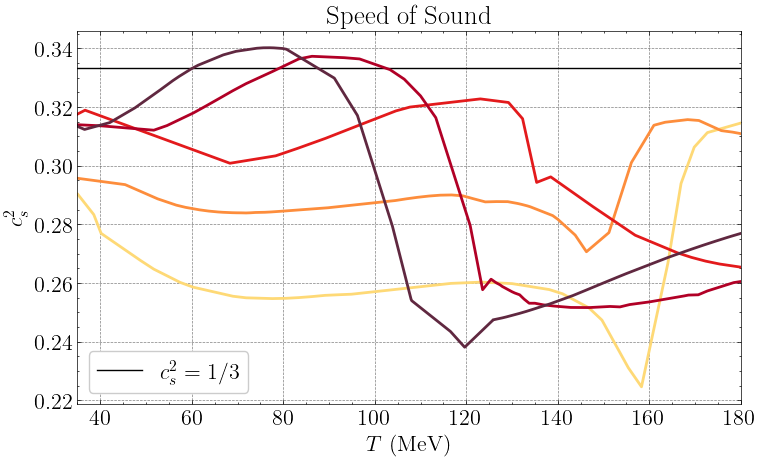}
        \label{fig:speed_of_sound}
  \end{minipage}
    \caption{ Trace anomaly (left) and speed of sound (right) as functions of the temperature for different lepton asymmetries. Solid lines correspond to the total trace anomaly, and dashed lines correspond to the QCD sector only. For high values of $|l_e + l_\mu|$, the trace anomaly reaches negative values inside the pion condensation phase and the speed of sound exceeds the conformal value.}
    \label{fig:Trace_anomaly_and_speed_of_sound}
\end{figure}

Finally, in Fig.~\ref{fig:Trace_anomaly_and_speed_of_sound} we show results for the trace anomaly $\Delta= (\epsilon - 3P)/T^4$ and speed of sound squared $c^2_s=d P/d \epsilon$ for different values of $|l_e + l_\mu|$. The trace anomaly results are similar to those of Ref.~\cite{Vovchenko2020crk}, except for the jumps at $T \approx 125 $~MeV caused by the first-order phase transitions. For large $|l_e + l_\mu|$, high values of charge and lepton chemical potentials are allowed, which makes the lepton contribution to the trace anomaly very large. The dashed lines show the result for the QCD sector only. Moreover, we notice that, for the interval $0.3 < |l_e + l_\mu| < 0.4$, the trace anomaly assumes a negative value during part of the trajectory. This region of negative values for the trace anomaly is reflected in the speed of sound exceeding the conformal value. 

\section{Discussion}\label{section: discussion}
As discussed in the previous section, for sufficiently high values of $|l_e + l_\mu|$, the universe may enter the pion condensation phase. Moreover, we have argued that this could happen through a first-order PT, which can lead to relevant consequences such as the production of primordial gravitational waves. In what follows, we briefly discuss the compatibility of the lepton asymmetries scenarios assumed here with observations and its implications for the possibility of having pion condensation in the early universe. Additionally, we comment on the possible production of primordial gravitational waves.
.
\subsection{BBN favored directions in the lepton asymmetry parameter space}

In Ref. \cite{Domcke:2025lzg} the authors attempted to establish an upper bound on the values of the primordial lepton asymmetries. They propose that such a bound depends on the flavor decomposition of the primordial asymmetries and on the choice of hierarchy of neutrino masses \cite{Lesgourgues:2006nd, Jimenez:2022dkn}. Neutrino mass eigenstates $m_1, m_2, m_3$ have two independent mass-squared differences and two possible hierarchies. If one takes $m_1 < m_2$ and the difference between $m_2^2$ and $m_1^2$ as the smaller mass-squared difference, then we call normal hierarchy (NH) the order $m_1 \leq m_2 \leq m_3$ and inverted hierarchy (IH) the order $m_3 \leq m_1 \leq m_2 $. With these considerations, the authors of Ref. \cite{Domcke:2025lzg} found the following directions on the lepton asymmetries parameter space that are consistent with CMB and favored by BBN data:
\begin{equation}
\begin{aligned}
        \textbf{(I).} \quad & l_e=0, \quad l_\mu =-l_\tau, \quad \text{(NH, IH)}\\
        \textbf{(II).}\quad & l_e= (-2/3)l_\mu, \quad l_\tau=(-1/3)l_\mu, \quad \text{(NH)} \\
        \textbf{(III).} \quad & l_e= - l_\mu, \quad l_\tau=0, \quad \text{(IH)}
\end{aligned}
\end{equation}

As discussed in Ref. \cite{Vovchenko2020crk} and confirmed within our approach, the possibility of the universe going through a pion condensation phase depends essentially on the value of the sum $|l_e + l_\mu|$. With this in mind, in direction \textbf{(I)} we expect the trajectories to cross the pion condensed phase when $l_\mu \geq 0.1$. Similarly, in direction \textbf{(II)} the condition is $l_\mu \geq 0.3$. Following the same reasoning, we expect the ``tau-phobic'' scenario, direction \textbf{(III)}, to be incompatible with the universe going through a pion condensation phase, since it implies $|l_e + l_\mu| = 0$.

\subsection{Gravitational Waves}
 At the early times considered here, photons are still coupled to the cosmic thermal bath. Therefore, their observation today cannot give us information about this early state of the universe. However, if a first-order PT occurred at this epoch, gravitational waves may have been produced via the collision of the growing bubbles of the true vacuum \cite{Caprini:2010xv, Kosowsky:1991ua} and via the induced turbulence and magnetic fields \cite{Hogan:1983zz}. As we have discussed in the previous sections, for $|l_e + l_\mu| \gtrsim 0.1 $ the quark-meson model predicts that the universe will undergo a first-order PT when entering the pion condensation phase, which could lead to an observable GW signal today.

Since the transitions discussed here occur at $T \sim 150$ MeV we may expect that the peak frequency is in a similar range as the GWs proposed to be generated during the QCD quark-hadron transition (in scenarios in which it is of first-order). Refs. \cite{Schwarz:1997gv, Schettler:2010dp, Caprini:2010xv} have found it to be $f_{\text{peak}} \sim 10^{-7}-10^{-9}$ Hz. Moreover first-order PTs produce a characteristic step in the power spectrum of the GW background \cite{Schwarz:1997gv}.

The latent heat $L$ per unit volume released at the first order phase transition for $|l_e + l_\mu|=0.2$, which is in the middle of the range considered here, is $L_2 \approx 2000$ MeV fm$^{-3}$. For $|l_e + l_\mu|=0.3$ and $|l_e + l_\mu|=0.4$ it can get three or four times larger, respectively. We can compare this to what is found in a simple bag model of the quark-hadron phase transition, in which $L=4B$, with $B$ the bag constant, ranging from $57\ \text{MeV/fm}^3$ to $500\ \text{MeV/fm}^3$ \cite{Schmid:1998mx}. A larger Latent heat generally leads to a stronger gravitational wave signal, since more energy will be available to be converted into GWs \cite{Kehayias:2009tn, Sagunski:2012lzm}. Therefore, a larger lepton asymmetry imbalance would result in a stronger GW signal. Moreover, the QCD quark-hadron transition and the pion condensation first-order transition could be distinguished by their different number of effective degrees of freedom, which affect the GW power spectrum.


The signal could, in principle, be detected by Pulsar Timing Arrays (PTAs) such as the Square Kilometre Array (SKA)~\cite{Caprini:2010xv, Weltman:2018zrl}. As pointed out in~\cite{Ghosh:2023aum}, a first-order phase transition during the QCD epoch can produce GWs in the nanohertz range and may account for the NANOGrav signal~\cite{NANOGrav:2023gor}, depending on the ratio \( \beta/H \) between bubble nucleation efficiency and the expansion rate. Therefore, the scenario proposed here could be a possible explanation for the signal. A detailed analysis of GW generation from pion condensation, its distinction from the quark-hadron transition, and its compatibility with the NANOGrav observation is left for a future work.

Note also that in case (B) (see Fig.~\ref{fig:cartoon_pion_condensation}), where both transitions along the cosmic trajectory are second-order, no additional gravitational waves are produced. However, an imprint can still be left on the preexisting GW power spectrum, which may have been generated during inflation \cite{Mukhanov:1990me}. The finite imbalance in the lepton asymmetries can increase the amplitude of the spectrum, and the change in the equation of state caused by the pion condensate can lead to an enhancement of the primordial GWs \cite{Vovchenko2020crk, hajkarim2019a}.

\subsection{Kaon condensation}
Due to the high charge chemical potentials reached for high lepton asymmetries, one could wonder about the formation of a kaon condensate. In this regard, we notice that results from chiral perturbation theory indicate that kaon and pion condensation cannot occur simultaneously \cite{Kogut:2001id, Mannarelli:2019hgn}. Moreover, a recent analysis, within the NJL model, has suggested that in the early universe pion condensation is always favored with respect to kaon condensation\cite{Cao:2024fyk}.

\section{Summary and outlook}\label{section: conclusions}

We used the quark-meson model as an effective description of QCD at finite $T$, $\mu_Q$ and $\mu_B$, and investigated the consequences of its phase diagram structure on the possible cosmic trajectories of the universe during the QCD epoch. We showed that, given large enough lepton asymmetries, $l_e + l_\mu \gtrsim 0.1$, the universe may enter the pion condensation phase through a first-order phase transition and leave it through a second-order phase transition. This corresponds to a new possible source of primordial GWs during the QCD epoch which might, in principle, be observed by PTAs. 

The phase transitions impact the pion and sigma condensates, the trace anomaly and the speed of sound along the cosmic trajectories. For large enough $l_e + l_\mu$ the trace anomaly may reach negative values, which corresponds to the speed of sound going above the conformal value. We also discussed the consistency of the occurrence of pion condensation with BBN and CMB data and noticed that it is only inconsistent with the ``tau-phobic'' scenario.

Pion condensation in the early universe could also generate other signatures through the effects of the equation of state on, e.g., the relic density of WIMP dark matter \cite{Hindmarsh:2005ix, Drees:2015exa} and the spectrum of primordial black holes \cite{Vovchenko2020crk, Bodeker:2020stj}. Primordial first-order phase transitions are also known to generate primordial magnetic fields \cite{Witten:1984rs, Cheng:1994yr} in addition to GWs. Finally, an intriguing possibility is that a first-order phase transition when entering the pion condensation phase could generate pion stars \cite{Carignano:2016lxe, Brandt:2018bwq} during the nucleation process.

Several improvements can be implemented in the future. First, the two-flavor model could be extended to three flavors, and one could include the Polyakov loop in the analysis. Although we do not expect any of these two extensions to modify the qualitative results, they could improve, for example, the quantitative agreement with the Lattice at $\mu_B=0$. Moreover, the inclusion of a Polyakov loop would allow for a discussion of both pion condensation and the deconfinement transition within the same model. Second, one can incorporate vacuum corrections that were neglected here and might affect the phase diagram, as discussed previously. Third, a detailed investigation of the process of generation of primordial GWs when entering the pion condensation phase would be interesting. In particular, discussing its distinguishability from the scenarios in which the QCD quark-hadron phase transition is of first-order and its potential as an explanation for the NANOGrav observation \cite{NANOGrav:2023gor}.

\begin{acknowledgments}
We would like to thank Bastian Brandt, Andreas Schmitt, Lorenz von Smekal, Dominik Schwarz and Dietrich Bodeker for helpful discussions. This work was partially supported by INCT-FNA (Process No. 464898/2014-5), CAPES (Finance Code 001), CNPq, and FAPERJ.
OF and JSB acknowledges support by the Deutsche Forschungsgemeinschaft (DFG, German Research Foundation) through the CRC-TR 211 'Strong-interaction matter under extreme conditions'– project number 315477589 – TRR 211. 
M.H. was supported in part by the Universidade Estadual do Rio de Janeiro through the Programa de Apoio à Docência (PAPD), and by the Brazilian National Council for Scientific and Technological Development (CNPq) under process No. 313638/2025-0.
OF is grateful for the kind hospitality of
 the astrophysics group at Goethe University Frankfurt, where this work was initiated.
\end{acknowledgments}

\appendix

\bibliographystyle{apsrev4-1}
\bibliography{references.bib}

\end{document}